\documentclass[twocolumn]{aa}
\usepackage{graphicx}
\usepackage{txfonts}
\usepackage{natbib}
\bibpunct{(}{)}{;}{a}{}{,}
\begin{document}
   \title{Binary formation within globular clusters : X-ray clues}

   \author{B. Gendre
          \inst{1}$^,$\inst{2}
          }

   \offprints{B. Gendre}

   \institute{
             CESR, 
	     9 avenue du Colonel Roche, 31028 Toulouse CEDEX 4 FRANCE\\
	\and
	      Present address : IASF-Roma,
              Via fosso del cavaliere 100, 00133 Roma ITALY\\
              \email{gendre@rm.iasf.cnr.it}
             }
   \date{Received ---; accepted ---}

   \abstract{We have investigated the effect of the number of primordial binaries on the relationship between the total number of detected binaries within globular cluster and its collision rate. We have used simulated populations of binary stars in globular clusters : primordial binaries and binaries formed through gravitational interactions. We show that the initial number of primordial binaries influences the relationship between the number of detected sources and the collision rate, which we find to be a power law. We also show that observing an incomplete sample provides the same results as those obtained with a complete sample. We use observations made by XMM-Newton and Chandra to constrain the formation mechanism of sources with X-ray luminosities larger than $10^{31}$ erg s$^{-1}$, and show that some of the cataclysmic variables within globular clusters should be primordial objects. We point out a possibly hidden population of neutron stars within high mass globular clusters with a low collision rate.
   }
   \maketitle

%


\section{Introduction}

Globular clusters harbor many X-ray sources \citep[see for example the review of ROSAT observations by ][]{ver01}. Some of these sources were found to be foreground or background objects not related to the clusters \citep[e.g. ][]{ver00}, but others are clearly associated with the clusters \citep[e.g.][]{gri01}. Using the latest X-ray observation facilities, it has been shown through X-ray spectroscopy \citep[see e.g. ][]{rut02, gen03b}, and by optical follow-up observations \citep[for example ][]{gri01, gri01b, poo02, web04} that the X-ray sources associated with the globular clusters are a mix of different binaries (cataclysmic variables, active binaries, low mass X-ray binaries) and millisecond pulsars \citep[see ][]{ver04}.

\citet{hen61} pointed out that a globular cluster made up of only single stars is unstable, and it will collapse in a few relaxation times. The relaxation time is defined as the time when the exchange of energy between cluster stars due to close encounters becomes significant for the cluster structure \citep[see the review of ][]{mey97}. Because this time is short  \citep[a typical value for a galactic globular cluster is a few hundred million years,][]{djo93} compared to the globular cluster ages and because we observe many clusters that have not yet collapsed, several mechanisms have been suggested to explain this : binary interactions, mass evaporation and black hole formation \citep{els87, hut92}. As pointed out by \citet{els87}, the mass evaporation should be most efficient at early times in the history of the cluster. Also, there is no firm evidence of a black hole in galactic globular clusters \citep{ver04}. On the other hand, simulations have shown that binaries can prevent the cluster core from collapsing \citep[see e.g. ][]{bau02}, leading to a stable configuration of the cluster, even if there is only a small sample of binaries within the cluster. This is due to the fact that binaries that interact with a cluster star can increase their binding energy and provide the star with the excess kinetic energy \citep{heg76}. Thus, binaries can help to keep the cluster stable \citep{hut92}, and binary interactions within globular clusters is a key point in studying the dynamical stability of globular clusters. Because most X-ray sources associated with globular clusters are binaries (or their offspring), X-ray observations of globular clusters can provide strong constraints on binary formation. Here, we study these X-ray clues using the latest observations made by Chandra and XMM-Newton.

This paper is organized as follows. In Sec.~\ref{sec_theo}, we briefly discuss the different ways a binary can form, and we explain the encounter rate within a cluster. In Sec.~\ref{sec_obs}, we review the observations made to date with Chandra and XMM-Newton as well as their main results. In Sec.~\ref{sec_simu}, we study, from a theoretical point of view, what should be expected from the observations, using simulated samples. We compare the results from the simulated samples to the observations in Sec.~\ref{sec_discu}.
\section{The encounter rate}
\label{sec_theo}
Binaries can form at the birth of a globular cluster (primordial binaries), or later, by dynamical processes (dynamical binaries) \citep[see ][ for a review about binaries within globular clusters]{hut92}. The number of dynamical binaries should scale with a parameter called the encounter rate \citep{cla75, fab75, ver87, ver03}.

According to \citet{ver03}, the encounter rate is :
\begin{equation}
\label{eq1}
\Gamma = \int A_{C,x} n_C n_x dV
\end{equation}

In this equation, $A_{C,x}$ represents the cross section of an interaction between species C and x, and $n_i$ represent the density of the species i. This value can be simplified with assumptions about the symmetry, equilibrium and gravitational potential of the cluster. This gives the expression (see Verbunt 2003 and Hut \& Verbunt 1983 for details):

\begin{equation}
\label{eq2}
\Gamma \propto \rho_0^{1.5} r_c^2
\end{equation}

The encounter rate can thus be calculated using only the core radius value, $r_c$, and the cluster central stellar density value $\rho_0$. If one of the hypotheses made in order to simplify this rate is false, then the encounter rate cannot be calculated with Eqn.~\ref{eq2} for this cluster. $\omega$ Cen provides a good example of such a limitation. Many sources in this cluster are detected outside the inner part of the cluster \citep{gen03a}. This fact conflicts with the mass segregation: the most massive sources (such as binaries) should gravitate toward the center of globular clusters \citep{mey97}. The equilibrium state of a cluster implies that mass segregation has taken place \citep{mey97}. Because mass segregation is not achieved in this cluster, the condition of equilibrium is not valid. 

These conditions imply use of this approximation only for globular clusters which are not collapsed \citep[the core radius is not defined in core-collapsed clusters,][]{kin62}. For these collapsed clusters, a more accurate method, used by e.g. \citet{poo03}, is to compute numerically the integration of the Eqn. \ref{eq1}.

In order to arrive at Eqn.~\ref{eq2}, \citet{ver03} assumes:
\begin{equation}
\label{eq3}
n_C = C_C \rho_0
\end{equation}

In Eqn.~\ref{eq3}, $C_C$ is the compact object density fraction, compared to the cluster central density. This parameter $C_C$ needs to be constant within the cluster (or at least within its central region, where the integration of Eqn.~\ref{eq1} is carried out). However, comparing several encounter rates calculated using Eqn. \ref{eq2} requires us to suppose that $C_C$ has the same value {\it in each cluster of the sample}. If one calculates the encounter rate using Eqn. \ref{eq2} and assumes that $\Gamma = \rho_0^{1.5} r_c^2$, care should be taken when using this hypothesis. 

We will now focus on the X-ray observations of globular clusters, and compare the results obtained with several different encounter rate calculation methods.

\section{X-ray observations}
\label{sec_obs}

With the new observatories XMM-Newton and Chandra, the detection and identification of X-ray sources related to globular clusters is now common \citep[see e.g.][]{gen03a,poo02}. This has lead to the study of the binary number versus the encounter rate by several authors \citep{gen03b, poo03,hei03}. However, there is still the problem of the completeness of binary samples (most of the faint binaries, such as active binaries have still not been detected in each cluster).  Thus these authors studied either the number of quiescent low mass X-ray binaries with a neutron star primary within globular clusters or the whole sample of detected binaries, without any assumption about their nature. 
\subsection{Low mass X-ray binaries within globular clusters}

Sources identified as a quiescent low mass X-ray binary with a neutron star primary (hereafter qLMXB$^{NS}$) were classified as such objects due to their soft spectrum and a luminosity of L$_x \sim 10^{32}$ erg s$^{-1}$ \citep[see e.g.][]{rut02}. These objects with a soft spectrum define the sample used by \citet{gen03b} and \citet{poo03}. We will now call them qNS. Due to their properties (soft sources with an X-ray luminosity around 10$^{32}$ erg s$^{-1}$), qNSs are fairly easy to detect in each cluster observed by XMM-Newton or Chandra. These sources were thus thought to form a complete sample of objects. Using the XMM-Newton and Chandra observations, \citet{gen03b}, \citet{poo03} and \citet{hei03} claimed that qNSs within globular clusters were formed by stellar encounters, and are thus not primordial binaries.

Nevertheless, another kind of qLMXB$^{NS}$ was detected by \citet{wij04}. These objects have a power law spectra not compatible with the soft spectrum of qNSs. According to \citet{wij04}, these objects cannot be securely classified using their X-ray color only because cataclysmic variables have similar spectra. Thus, some sources classified as cataclysmic variables could be in fact qLMXBs \citep{wij04}, and the qLMXB$^{NS}$ sample should not be considered to be complete.

\subsection{Sources detected within globular clusters}

Using numerical integration of Eqn. \ref{eq1}, \citet{poo03} indicated that their whole sample of globular cluster X-ray sources was compatible with a power law relationship between the number of binaries and the encounter rate.Using the encounter rate calculated with the simplified Eqn. \ref{eq2}, and another sample which meets the conditions required by using Eqn. \ref{eq2}, \citet{gen04} confirmed this result. \citet{hei03} used  another method and fitted their sample with a law following the $r_c^{\alpha}\rho_0^{\beta}$ relationship ($\alpha$ and $\beta$ left free during the fit). Because the encounter rate is given by Eqn. \ref{eq1}, this method implies the same limitations as the ones indicated in deriving Eqn. \ref{eq2}. Their best fit values were respectively 1.0 and 1.3, and their sample excluded the values of 1.5 and 2.0 \citep{hei03}. One should note that :

\begin{equation}
\rho_0^1 r_c^{1.3} = (\rho_0^{1.5} r_c^2)^{2/3}
\end{equation}

Thus, one can understand the \citet{hei03} work as a claim of a power law relationship between the number of X-ray binaries and the encounter rate given by Eqn. \ref{eq2}. \citet{hei03} have also studied the effect of the luminosity limit variation or the spectral characteristics of their sample. They again found a correlation which can be interpreted as a power law relationship. Moreover, they report a change in their best fitted values that can be interpreted as a change in the power law index. 

 Three different groups, using different methods of collision rate calculation, have thus obtained the same result : the number of detected sources located within globular clusters scales as a power law of the cluster encounter rate. Each group results are independent : \citet{gen04} does not use the result and the encounter rate values from \citet{poo03} or \citet{hei03}, and neither \citet{poo03} or \citet{hei03} use the results from other groups. There is thus no problem when one compares these results {\it as long as two different samples (from two different groups) are not merged without recalculating the encounter rate in a common manner}.

 In this paper, we compare the relationship between the source number and the encounter rate. In Sec. \ref{sec_simu} we will use the encounter rate as a number, and we will not focus on how this number is obtained (supposed valid). In Sec. \ref{sec_discu} we will use the sample of \citet{gen03b} and thus his collision rate calculation method to discuss the number of qLMXB$^{NS}$ within globular clusters.

One should then explain why a linear relationship, observed for the qNS sample, appears to become a power law relationship for other sources. There are three possible explanations for such a behavior. First, the hypotheses made for the derivation of the collision rate are not valid for all the objects.  Second, this could be due to the sample completeness. Finally, there could be a fraction of primordial binaries within the sample. We have examined the last two possibilities using simulated samples.

\section{Simulations}
\label{sec_simu}
   \begin{table}[b!]
      \caption[]{Simulation of the effect of completeness. We have simulated two distributions of sources which obey the linear relationship given in Eqn.~\ref{eq2}.  We have then excluded a part of this sample and looked for a  relationship. The power law index is indicated in this table together with its error (errors are quoted at the 1$\sigma$ level).}
         \label{table_completude}
     $$ 
         \begin{tabular}{ccc}
            \hline
            \noalign{\smallskip}
            Sample     &   Gaussian & Flat\\
	     excluded  &   distribution& distribution\\
            \noalign{\smallskip}
            \hline
            \noalign{\smallskip}
        5 \%  & 1.02 $\pm$ 0.03 & 1.02 $\pm$ 0.03\\
        10 \% & 1.02 $\pm$ 0.05 & 1.02 $\pm$ 0.05\\
        25 \% & 1.02 $\pm$ 0.06 & 1.03 $\pm$ 0.06\\
	50 \% & 1.00 $\pm$ 0.08 & 1.01 $\pm$ 0.09\\
	75 \% & 0.91 $\pm$ 0.12 & 0.92 $\pm$ 0.12\\
            \noalign{\smallskip}
            \hline
         \end{tabular}
     $$ 
   \end{table}

We have found a linear relationship for a sample that we thought to be complete, but a power law relationship for an incomplete sample. The most natural explanation thus involves the completeness of the sample. To test this, we have used a simulation of a globular cluster sample. We have chosen a sample of 33 clusters. Their encounter rates are fixed in order to be regularly spaced between two limiting values (0 and 100 using the conventions of Gendre et al. 2003b). First, we populate our sample with binaries using the linear relationship between the number of sources and the collision rate. We have chosen the number of binaries in a given cluster to be $0.5 \times \Gamma$. 

We then assigned a luminosity to each binary in each cluster. We can use two different distributions: a Gaussian-like (i.e. there is a 'most probable' luminosity), and a flat distribution (all luminosities have the same probability). These distributions were chosen to simulate two observational cases. The Gaussian distribution can deal with a rising luminosity function which is cut off at low luminosities by the instrument sensitivity. We have chosen the Gaussian parameters to be $4.0 \times 10^{31}$ (mean) and $1.0 \times 10^{31}$ (standard deviation) erg.s$^{-1}$. The flat distribution deals with a unique population of object, where one should expect that each luminosity has the same probability of appearing. The boundaries of this distribution have been chosen to be  $10^{30}$ and $10^{33}$ erg.s$^{-1}$.

To test the effect of completeness, we then excluded a part of the sample. Because we are dealing with an observational bias, we have excluded the faintest binaries of the sample. We have chosen to exclude 5, 10, 25, 50 or 75 \% of the binary population. The flat and Gaussian distribution parameters are chosen arbitrarily, but because we exclude a percentage of the binary population, we can rescale the distribution parameters to any realistic parameters without changing the simulation results. We then tried to find a power-law relationship between the collision rate and the number of remaining objects. The simulation is done 500000 times for both luminosity distributions with a given threshold of binary exclusion (thus, a total of 5 000 000 different simulations). At the end of each simulation, we compute the value of the power-law relationship index. At the end of all simulations, we compute the mean values of the indexes, which are listed in Table~\ref{table_completude}. For both distributions, the power law index calculated is always compatible with unity.

 We have also looked for the effect of adding to the number of dynamical binaries a random number value. This random number was chosen according to a Gaussian distribution. We then looked for a power-law relationship between the encounter rate and this number of dynamical binaries added to a random value. The simulation is again repeated 500000 times for each of the Gaussian parameters listed in Table \ref{table_simu}. The results of these simulations are given in Table~\ref{table_simu}, in the 'population 1' column.

 The number of primordial binaries should scale with the initial parameters of the cluster \citep{mur91}. Because the encounter rate is based on current parameters, and because the evolution of the initial parameters depends also on parameters external to the cluster (e.g. the galactic field), we should not expect the current encounter rate to be linked to the initial parameters. Because of this fact, they could be uncorrelated. One can then assume them (and thus the number of primordial binaries) to be random numbers within the collision rate parameter space. Thus, the random number added to the number of dynamical binaries could represent the primordial binariy number.

 The assumed distribution of the primordial binariy number (a Gaussian distribution) is somewhat arbitrary. Ideally, this distribution should take into account the evolution effects, the fact that not all primordial binaries are X-ray sources, and the fact that we do not detect all X-ray sources due to the limiting luminosity. Because most of these effects are now only partly understood, we have assumed that there is a {\it most probable number of primordial binaries} and modified the Gaussian standard deviation rather than testing other distributions (such as a flat or a cut-off power law distribution). We have tried to estimate the evolution effects, because gravitational interactions can destroy primordial objects \citep{poo02b,ver03}.

 A primordial object is ``destroyed'' if it is disrupted or modified by a dynamical encounter. Such an event occurs during an exchange encounter \citep[using the formulation of ][]{ver03}. According to \citet{ver03} the exchange encounter rate is similar to the collision rate given in Eqn. \ref{eq1}. The differences are that we now talk about the binary density (not the compact object density) and that the cross section depends on the binary size (not the stellar size). The size of a binary is larger than that of a star \citep[$\sim$3 stellar radii, ][]{ver03}; on the other hand, the number of primordial X-ray binaries (considered here, so $\sim$5 within each cluster) is less than the number of known compact objects within some globular cluster \citep[see e.g. the case of 47 Tuc, ][]{gri01}, which is certainly less than the number of compact objects really present within these clusters (see Sec. \ref{sec_discu}). Because we do not know precisely all of these parameters, we have decided to use the same relationship to fix the number of primordial binaries dynamically destroyed and the number of dynamical binaries. We compared the number of primordial binaries which are present within the cluster (given by the random number generator) and the number of primordial binaries destroyed within the cluster. If there are more primordial binaries destroyed than the number of primordial binaries present in the cluster, the total number of binaries is fixed to the value of dynamical binaries only. Otherwise, the total number of binaries is the sum of the number of remaining primordial binaries and the number of dynamical binaries.  The search for a power-law relationship is finally repeated, and this simulation is again done 500000 times. Table~\ref{table_simu} gives the results of this simulation in the 'population 2' column. 

 From Table \ref{table_simu}, one can see that the power law index is no longer compatible with unity for either population 1 or population 2. The variation of the proportion of primordial binaries, simulated by varying the parameters of the Gaussian distribution of the random number, always disagrees with a linear relationship. Moreover, as one can see in Table~\ref{table_simu}, the power law index decreases when the primordial binary fraction increases. These results are discussed in the next section.

   \begin{table}[t!]
     \caption[]{Simulation of the effect of the presence of primordial binaries on a sample of dynamically formed binaries. We have simulated two distributions containing a mixture of sources that obey the linear relationship given in Eqn.~\ref{eq2} and a random number which simulates the primordial source number. We have searched for a power law relationship within this distribution.  Population 1 is a population where there is no dynamical destruction, while population 2 is a population where primordial binaries can be destroyed by dynamical interactions. The power law index is indicated in this table together with its error (errors are quoted at the 1$\sigma$ level). The Gaussian parameters (in unit of sources) are the parameters of the distribution of the random number simulating the primordial binary number.}
         \label{table_simu}
     $$ 
         \begin{tabular}{cccc}
            \hline
            \noalign{\smallskip}
            Gaussian     &   Gaussian & Population 1&Population 2\\
	     mean  &   sigma & &\\
            \noalign{\smallskip}
            \hline
            \noalign{\smallskip}
         5  & 3 & 0.63 $\pm$ 0.10 & 0.72 $\pm$ 0.10\\
        10  & 3 & 0.48 $\pm$ 0.06 & 0.50 $\pm$ 0.06\\
         5  & 5 & 0.64 $\pm$ 0.12 & 0.70 $\pm$ 0.13\\
	10  & 5 & 0.49 $\pm$ 0.09 & 0.51 $\pm$ 0.09\\
	20  & 10& 0.36 $\pm$ 0.10 & 0.29 $\pm$ 0.10\\
            \noalign{\smallskip}
            \hline
         \end{tabular}
     $$ 
   \end{table}

\section{Discussion}
\label{sec_discu}
\subsection{The qNS sample}
As already indicated, qNSs were thought to form a complete sample, and their number follow a linear relationship with the encounter rate \citep{gen03b, poo03, hei03}. If we assume that this relationship is valid, then we also need to assume that the encounter rate calculated by \citet{gen03b} is valid, and thus that $C_c$ is similar within each cluster. In other words, the neutron star density is constant within each cluster (the more massive a cluster, the more numerous the neutron stars are). 

 Neutron stars can be detected as millisecond pulsars. As not all the millisecond pulsars within a globular cluster can be detected \citep[due to limiting sensitivity and the small opening of the cone angle, ][]{ran03a, ran03b}, the number of known neutron stars within one cluster should be taken as a lower limit to the real number of neutron stars within this cluster. The hypothesis that the neutron star density is constant cannot thus be verified. If this hypothesis is true, then a hidden population of neutron stars is present, at least in massive clusters with no known neutron stars (such as M22). This population could be composed of millisecond pulsars which have not yet been detected, but it could also consist of isolated cold neutron stars (which cannot emit X-ray light) or double neutron star binaries (which could be detected from their gravitational waves).

We can also assume that $C_c$ is not similar within each cluster, but then the relationship \citet{gen03b} found implies that this parameter cannot vary too widely, otherwise the collision rate might vary too widely from the computed values (and thus disrupt the relationship). We thus need to assume, again, a hidden population of neutron stars (despite the fact that this population may be smaller than if the neutron star density did not vary at all).

One may argue that the sample to consider would be the whole qLMXB$^{NS}$ sample \citep[qNSs, plus the 'hard' qLMXB$^{NS}$ indicated by ][]{wij04} rather than the qNS one. Because our current sample is then not complete, we cannot draw any conclusions.  The 'hard' qLMXB$^{NS}$ EXO1745-248 have very different spectral properties than the qNS : \citet{wij04} have shown that the canonical spectral model for soft qNS \citep[a neutron star atmosphere model, ][]{pav92, zav97} cannot fit its spectrum. This may be due to a difference in the system properties, and we cannot rule out that 'hard' qLMXB$^{NS}$ and qNS may be two distinct kinds of qLMXB$^{NS}$, which could form in different ways. We also cannot rule out that these objects are the same kind of qLMXB, and that indeed our sample is not complete. But our simulations (see Table \ref{table_completude}) indicate that the non completeness of a sample does not change the correlation results or the value of the power law relationship (it only increases the error bars).

\subsection{The primordial binaries versus the dynamically formed binaries}
Our simulations have shown that there is no effect caused by the incompleteness of a sample (see Table \ref{table_completude}). Thus, any incomplete sample would allow us to study the correlation between the number of objects and the encounter rate. Table~\ref{table_simu} clearly shows that a population which is a mixture of primordial binaries and dynamically formed binaries has a correlation index which is less than one. The observations made by XMM-Newton and Chandra clearly indicate that the power law index of the correlation between the number of X-ray sources within a cluster and its collision rate is less than one. \citet{poo02} have found 0.74 $\pm$ 0.36, \citet{gen04} found 0.5 $\pm$ 0.2 and \citet{hei03} have shown that the value of one is excluded at the 90 \% confidence level. This may indicate that the X-ray sources associated with the globular clusters are a mixture of primordial binaries and dynamically formed objects. The variations of the exposant indexes of the core radius and the central density, as reported by \citet{hei03} can be interpreted as a variation of the powerlaw index, and may then be due to a change in the luminosity limit or in the spectral properties of the sample which modifies the proportion of primordial binaries.

At a limiting luminosity of L$_x = 4\times 10^{30}$ erg s$^{-1}$ \citep[used by ][]{poo03}, the population is dominated by a mixture of qLMXB$^{NS}$ and cataclysmic variables. If we assume (see above) that the 'soft' qLMXB$^{NS}$ are formed by dynamical interactions, then the remaining objects ('hard' qLMXB$^{NS}$ and CVs) should (at least for some of them) be primordial objects. One should note from the study of \citet{hei03} that their hard and luminous sample (their Fig. 5) is more compatible with a linear relationship (it is excluded at the 50\% confidence level only) than their whole hard sample (their Fig. 6). This could indicate that the proportion of primordial binaries is less within the hard and luminous sample than in the hard whole sample. If so, this could be an indication of dynamical formation for some of these hard and luminous sources.

\section{Conclusions}

We have presented simulations of X-ray source populations and compare the simulated samples with the observations of XMM-Newton and Chandra in order to constrain the formation mechanisms of binaries within globular clusters. Using the whole sample of detected X-ray sources within globular clusters, we have presented evidence to support the idea that globular clusters contain primordial binaries, and that some cataclysmic variables should be primordial objects. The growing sample of X-ray sources identified and thus associated with such kinds of objects will allow us to refine the X-ray samples and thus to improve these results.

\begin{acknowledgements}
This work is an updated part of my PhD thesis, defended in Toulouse (France) in January 2004. I would like to thank D. Barret, my PhD supervisor, J.M. Hameury and J.P. Lasota, the two thesis referees, J. Ballet, M. Auriere and P. von Ballmoos, the three thesis examiners, for very useful comments; this updated version largely uses their advice. I would also like to thank N.A. Webb for comments and corrections of this text, and D. LeQueau and E. Jourdain for encouragement during the thesis writing. I finally thank the anonymous referee for a very constructive report and useful comments.

\end{acknowledgements}

\end{document}